\documentclass[prl,twocolumn,floatfix,superscriptaddress]{revtex4-2}
\usepackage{dcolumn,amsmath}
\usepackage{graphicx}
\usepackage{bm}
\usepackage{hyperref}
\usepackage{xcolor}

\hypersetup{colorlinks=true, citecolor=blue, urlcolor=blue, linkcolor=blue}

\usepackage[utf8]{inputenc}

\usepackage[T2A]{fontenc}

\newcommand{\cm}{\ensuremath{{\rm cm}}$^{-1}$}

\usepackage{tabularx}
\usepackage{braket}
\usepackage{xtab,afterpage}
\usepackage{multirow}
\usepackage{booktabs}

\usepackage{xcolor}

\begin{document}

\title{\textit{Ab initio} study of electronic states and radiative properties of the AcF molecule}

\author{Leonid~V.~Skripnikov}
\email{skripnikov\_lv@pnpi.nrcki.ru,\\ leonidos239@gmail.com}
\affiliation{Petersburg Nuclear Physics Institute named by B.P.\ Konstantinov of National Research Center ``Kurchatov Institute'' (NRC ``Kurchatov Institute'' -- PNPI), 1 Orlova roscha, Gatchina, 188300 Leningrad region, Russia}
\affiliation{Saint Petersburg State University, 7/9 Universitetskaya nab., St. Petersburg, 199034 Russia}

\author{Alexander~V.~Oleynichenko}
\email{oleynichenko\_av@pnpi.nrcki.ru, \\ alexvoleynichenko@gmail.com}
\affiliation{Petersburg Nuclear Physics Institute named by B.P.\ Konstantinov of National Research Center ``Kurchatov Institute'' (NRC ``Kurchatov Institute'' -- PNPI), 1 Orlova roscha, Gatchina, 188300 Leningrad region, Russia}
\homepage{http://www.qchem.pnpi.spb.ru}

\author{Andr\'ei~Zaitsevskii}
\affiliation{Petersburg Nuclear Physics Institute named by B.P.\ Konstantinov of National Research Center ``Kurchatov Institute'' (NRC ``Kurchatov Institute'' -- PNPI), 1 Orlova roscha, Gatchina, 188300 Leningrad region, Russia}
\affiliation{Department of Chemistry, M.V. Lomonosov Moscow State University, Leninskie gory 1/3, Moscow, 119991~Russia}

\author{Nikolai~S.~Mosyagin}
\affiliation{Petersburg Nuclear Physics Institute named by B.P.\ Konstantinov of National Research Center ``Kurchatov Institute'' (NRC ``Kurchatov Institute'' -- PNPI), 1 Orlova roscha, Gatchina, 188300 Leningrad region, Russia}
\homepage{http://www.qchem.pnpi.spb.ru}

\author{Michail~Athanasakis-Kaklamanakis}
\affiliation{Experimental Physics Department, CERN, CH -- 1211 Geneva 23, Switzerland}
\affiliation{KU Leuven, Instituut voor Kern- en Stralingsfysica, B -- 3001 Leuven, Belgium}

\author{Mia Au}
\affiliation{Systems Department, CERN, CH -- 1211 Geneva 23, Switzerland}

\author{Gerda Neyens}
\affiliation{KU Leuven, Instituut voor Kern- en Stralingsfysica, B -- 3001 Leuven, Belgium}

\begin{abstract}
Relativistic coupled-cluster calculations of the ionization potential, dissociation energy, and excited electronic states under 35,000~cm$^{-1}$ are presented for the actinium monofluoride (AcF) molecule. The ionization potential is calculated to be IP$_e=48,866$~\cm, and the ground state is confirmed to be a closed-shell singlet and thus strongly sensitive to the $\mathcal{T}$,$\mathcal{P}$-violating nuclear Schiff moment of the Ac nucleus. Radiative properties and transition dipole moments from the ground state are identified for several excited states, achieving an uncertainty of $\sim$450~\cm for the excitation energies. For higher-lying states that are not directly accessible from the ground state, possible two-step excitation pathways are proposed. The calculated branching ratios and Franck-Condon factors are used to investigate the suitability of AcF for direct laser cooling. The lifetime of the metastable $(1)^3\Delta_1$ state, which can be used in experimental searches of the electric dipole moment of the electron, is estimated to be of order 1~ms.
\end{abstract}

\maketitle

\section{Introduction}\label{sec:intro}

Radioactive molecules containing heavy atoms provide unique opportunities to search for new physics beyond the Standard model (SM)~\cite{ArrowsmithKron:23}. The yet undiscovered realm can manifest itself in interactions violating time-reversal ($\mathcal{T}$) and/or parity ($\mathcal{P}$) symmetries~\cite{Ginges:04,Graham:13,Roberts:15,Safronova:18,Alarcon:22}, and the shifts that such interactions induce in the energies of molecular electronic states can be probed in modern tabletop-scale experiments.

There are several key factors that make molecules of heavy-element compounds highly valuable for such experiments. Large internal effective electric fields and the presence of closely spaced levels of opposite parity greatly increase the sensitivity to symmetry-violating interactions (see Refs.~\cite{Khr91,Khriplovich:97,KL95,Safronova:18} and references therein). Furthermore, it is possible to design molecules with heavy atoms which can be laser-cooled, resulting in increased coherence times~\cite{Isaev:RaF:10,isaev2017laser,Kozyryev:17,Lim:18,Hao:BaF:19,Augenbraun:20,Oleynichenko:2021AcOH,Augenbraun:21,Zulch:22,Isaev:22}. Finally, heavy elements typically have a large number of isotopes with different nuclear properties, some of which have been predicted to possess outstandingly large $\mathcal{P}$- and $\mathcal{T,P}$-violating nuclear electromagnetic moments induced by yet unknown nucleon-nucleon interactions~\cite{Ginges:04}. The $\mathcal{P}$-violating nuclear anapole moment and the $\mathcal{T,P}$-violating nuclear Schiff and magnetic quadrupole moments are among them. Nuclear structure theory predicts that these moments can be greatly enhanced in octupole deformed nuclei
~\cite{Auerbach:96,Dobaczewski:05,Auerbach:06,Butler:20,Flambaum:Feldmeier:20,Flambaum:MQM:22}.

The direct observation of $\mathcal{T}$,$\mathcal{P}$-violating nuclear Schiff moments can help to shed light on the overwhelming imbalance of matter and antimatter in the Universe and the $\mathcal{CP}$ problem of quantum chromodynamics, both of which cannot be explained within the SM~\cite{Sakharov1967,MatterAntimatter:2003,Safronova:18}. Highly sensitive experiments attempting to make the first successful measurement of a nuclear Schiff moment across the nuclear chart can set new limits on the quantum chromodynamics vacuum angle $\bar{\theta}$ and other hadronic $\mathcal{CP}$-violation parameters. The most suitable species for such experiments are diamagnetic atoms and molecules. Thus, several experiments have been performed using Hg~\cite{Romalis:01,Griffith:09,Graner:16} and Xe~\cite{Rosenberry:01,Sachdeva:19,Allmendinger:19} atoms and the TlF molecule~\cite{Cho:91,Norrgard:17,Grasdijk:21}. Moreover, a solid-state experiment sensitive~\cite{Skripnikov:16a} to the oscillating nuclear Schiff moment of $^{207}$Pb has been recently performed~\cite{aybas2021}.

Despite the global efforts, no nuclear Schiff moment has been successfully measured so far, and only upper bounds have been placed by the various experiments. To make progress towards a successful measurement, nuclei with enhanced Schiff moments and compounds that are the most sensitive to these moments need to be identified. Recently, a series of diatomic molecules containing octupole-deformed lanthanide and actinide nuclei, namely, $^{227}$AcF, $^{227}$AcN, $^{227}$AcO$^+$, $^{229}$ThO, $^{153}$EuO$^+$ and $^{153}$EuN, was proposed~\cite{Flambaum:Dzuba:20,Skripnikov:2020c}. For these molecules, estimated energy shifts~\cite{Skripnikov:2020c} arising from the interaction between the nuclear Schiff moment and molecular electrons were predicted to be up to three orders of magnitude greater than in TlF, making these systems promising for the next generation of experimental campaigns. 
Since a strong enhancement in the nuclear Schiff moment is expected for the $^{225}$Ac and $^{227}$Ac isotopes~\cite{Verstraelen:19,Flambaum:Feldmeier:20,Dalton:23}, actinium compounds are highly promising candidates for such experiments. $^{225}$Ac and $^{227}$Ac are rather long-lived with $T_{1/2} = 10$ days and $T_{1/2} = 21.8$ years, respectively, and supplies are commercially available as $^{225}$Ac is investigated for targeted-$\alpha$ cancer therapy~\cite{Morgenstern2020}.

Radioactive nuclides can be produced at accelerator facilities such as ISOLDE at CERN~\cite{Catherall2017} and delivered in the form of radioactive molecular ion beams~\cite{Au2023}. Current production techniques involve extraction from a target material and subsequent ionization, and thus benefit from theoretical predictions of molecular properties such as the ionization potential and dissociation energy. Radioactive molecules containing isotopes with half-lives down to a few tens of milliseconds can be produced and studied, as was demonstrated recently with the study of several isotopologues of RaF~\cite{GarciaRuiz:20,Udrescu:21,Udrescu:23} at the CRIS (Collinear Resonance Ionization Spectroscopy) experiment~\cite{Cocolios:2016,Vernon:2020}. Both broadband and narrowband collinear laser spectroscopy can be performed, and similar studies are planned on AcF molecules with a multidisciplinary motivation of fundamental, nuclear, and medical physics~\cite{AcF:Proposal:21}.

No experimental data exist for AcF as of yet, while the available theoretical information is also quite modest. Relativistic coupled cluster calculations~\cite{Skripnikov:2020c} predict AcF to have a closed-shell ground state with an equilibrium internuclear distance of $r_e = $~2.12~\AA, and give a preliminary theoretical estimate of its ionization potential (IP), 48,600(250)~cm$^{-1}$~\cite{AcF:Proposal:21}. Excitation energies for transitions to six excited electronic states were predicted~\cite{AcF:Proposal:21}, but the picture is not complete since the two-electron excitations with respect to the electronic ground state were not accounted for in the employed theoretical model. Furthermore, no predictions for the transition dipole moments have been made so far, even though they are necessary to locate the strongest transitions and thus to optimally design the first spectroscopic search. Lastly, a theoretical estimate remains to be made regarding the suitability of AcF for direct laser cooling, since the isovalent TlF molecule is predicted to be laser-coolable~\cite{Hunter:TlF:12}.

Here, we expand the available information about AcF with a comprehensive theoretical study of its electronic structure. Firstly, we re-examine the value of the ionization potential and calculate the dissociation energy. Secondly, we present the general picture of the excited electronic states and their spectroscopic constants. Afterwards, the most promising one- and two-step electronic transitions from the ground state for future spectroscopic measurements are proposed on the basis of the calculated transition dipole moments. Finally, we discuss the laser-coolability of AcF and its possible use in future experiments aimed at searches of $\mathcal{T,P}$-violating interactions.

\section{Computational methods}\label{sec:comput}

\subsection{Ionization potential calculations}

The IP of AcF can be determined as the difference between the total energies of the neutral AcF molecule and the AcF$^+$ cation in their electronic ground states (IP$_e$), and also with taking into account the zero-vibrational contribution (IP$_0$). Since electronic ground-state wavefunctions of AcF and AcF$^+$ are dominated by a single Slater determinant, the relativistic single-reference coupled-cluster theory (RCC) is the optimal choice to calculate the IP. Within this approach, the exact electronic wave function $\Psi$ can be written in the exponential form (see e.g. Refs.~\cite{Visscher:96a,Bartlett:2007} and references therein for a comprehensive review of the CC theory):
\begin{equation} 
\label{expAnsatz}
    \ket{\Psi} = e^{\hat T} \ket{\Phi_0},\quad\quad {\hat T}=\hat{T}_1 + \hat{T}_2 + \hat{T}_3 + \dots
\end{equation}
where $\Phi_0$ is the reference determinant and the cluster operator $T$ consists of $n$-body terms representing single ($\hat{T}_1$), double ($\hat{T}_2$), triple ($\hat{T}_3$), etc. excitations with respect to $\Phi_0$. The basic CCSD model necessarily includes only single and double excitation operators ($\hat{T} \approx \hat{T}_1 + \hat{T}_2$), and various approximate schemes accounting for higher excitations have been developed.

The simplest and least computationally demanding approach to account for triple excitations is the CCSD(T) model, which involves the solution of the CCSD equations followed by the non-iterative calculation of the approximate fifth-order perturbation correction arising from connected triples~\cite{Raghavachari:89}. The more sophisticated CCSDT-3 model also estimates the triple-excitation amplitudes perturbatively, but they are now included into the iterative solution of equations for the $\hat{T}_1$ and $\hat{T}_2$ amplitudes~\cite{Noga:CCSDT-3:87}. Even better level of accuracy is attained within the full CCSDT model~\cite{Noga:CCSDT:87}, which treats the $\hat{T}_3$ operator in a fully consistent iterative manner without any additional neglections and approximations. Contributions from the quadruple-excitation operator $\hat{T}_4$ can be accounted for within corresponding similar approximations CCSDT(Q) and CCSDTQ-3~\cite{Bomble:05,Kallay:6}. It must be noted, however, that the computational cost grows very quickly when going from CCSD(T) to CCSDTQ-3.

To calculate the IP and the dissociation energy ($D_e$) of AcF, the following computational scheme was used. The leading-order value was obtained within the relativistic CCSD(T) approach using the Dirac-Coulomb Hamiltonian. Here, 70 electrons of AcF were included in the correlation treatment. The virtual molecular orbital energy cutoff was set to 300 Hartree. The MBas basis set was employed, corresponding to the extended uncontracted all-electron triple-zeta Dyall's AETZ~\cite{Dyall:07,Dyall:12} basis set for Ac and the uncontracted augmented all-electron quadruple-zeta AAEQZ~\cite{Dyall:04,Dyall:2010,Dyall:F:16} basis set for F.

Higher-order correlation effects were accounted for with a scheme of incremental corrections. The first correction was calculated as the difference between the IP ($D_e$) values calculated at the CCSDT-3 and CCSD(T) levels, with both calculations performed with 38 correlated electrons and the two-component generalized relativistic pseudopotential (GRPP) model~\cite{Titov:99,Mosyagin:10a,Mosyagin:16,Mosyagin:17} for the Hamiltonian. At this stage, the energy cutoff for virtual orbitals was set to 35~Hartree and a compact basis set (CBas) of atomic natural orbitals was generated using the procedure developed and described in Refs.~\cite{Skripnikov:2020e,Skripnikov:13a}. The basis set for Ac thus consisted of 8$s$-, 8$p$-, 7$d$-, 4$f$-, 4$g$- and 2$h$-type contracted functions and can be denoted as $(20s20p20d15f15g15h)/[8s8p7d4f4g2h]$, where the numbers in the $()$-brackets are the numbers of primitive Gaussian functions included in the contracted functions listed in the $[]$-brackets. The aug-cc-pVTZ-DK basis set~\cite{Dunning:89,Kendall:92,augCCPVXZDK} was used for F. 

The next correlation correction included the difference between the IP values obtained within the CCSDT and CCSDT-3 approaches. Due to the very high computational cost of the CCSDT model, only 28 electrons of AcF were correlated and the basis set used was further reduced to $(20s20p20d15f)/[8s8p7d4f]$ for Ac and aug-cc-pVDZ-DK set~\cite{Dunning:89,Kendall:92,augCCPVXZDK} for F; we call this basis set SBas. The virtual energy cutoff was set to 10 Hartree. To derive the contribution of connected quadruple-cluster amplitudes, we calculated the difference between IP values obtained within the CCSDT(Q) and CCSDT approaches using the similar parameters as in the ``CCSDT $-$ CCSDT-3'' correction, except that the virtual orbitals energy cutoff was set to 5 Hartree. The last correction for even higher-order correlation effects was obtained as the difference between the CCSDTQ-3 and CCSDT(Q) values, correlating the 18 outermost electrons. In this calculation, the virtual orbital energy cutoff was reduced to 4 Hartree and the Ac basis set was reduced to $(20s20p20d15f)/[8s6p5d3f]$. Similar corrections were also used to calculate $D_e$. 

To account for the incompleteness of the basis set, corrections to the IP and $D_e$ were calculated using the scalar-relativistic part of the GRPP operator. Such a simplification of the Hamiltonian allowed us to employ a very large basis set (LBas) consisting of $[30s30p30d30f15g15h15i]$ uncontracted basis functions for Ac and the extended Dyall's uncontracted AAEQZ~\cite{Dyall:F:16} set for F. In the latter case, we replaced the original basis functions of the AAEQZ basis with $l=3,4,5,6$ by corresponding functions of the aug-cc-pV7Z~\cite{Feller:1999} basis set. The basis set correction was thus calculated as the difference between the CCSD(T) values of the IP obtained using the LBas versus the MBas basis sets, including 38 AcF electrons in the correlation treatment.

The contribution of the Gaunt electron-electron interaction was estimated using the molecular-mean-field exact two-component approach~\cite{Sikkema:2009} at the Fock-space RCCSD (FS-RCCSD) level. Finally, the contribution of quantum electrodynamics (QED) effects was taken into account using the model QED Hamiltonian approach~\cite{Shabaev:13,Malyshev:2022} adapted for molecular calculations~\cite{Skripnikov:2021a}.

Relativistic calculations of the IP and $D_e$ were carried out using the {\sc dirac}~\cite{DIRAC_code:19,Saue:20} and {\sc mrcc}~\cite{MRCC:program:20,Kallay:MRCC:20,Kallay:1,Kallay:2} packages. Scalar relativistic correlation calculations were performed using the {\sc cfour}~\cite{CFOUR,Matthews:CFOUR:20} code. 

\subsection{Excited-state calculations}

Qualitatively, the electronic states of AcF can be regarded as the states of Ac$^+$ split by the field of the F$^{-}$ anion. The atomic spectrum of Ac$^+$ is dense, with several dozens of states observed below $\sim$50,000~cm$^{-1}$~\cite{Sansonetti:05}. In the energy range below 30,000~cm$^{-1}$, the Ac$^+$ states are dominated by the $7s^2$, $7s6d$, $6d^2$, $7s7p$, and $7p6d$ configurations, while for higher-lying states the $7s5f$, $7s8s$, and $6d5f$ configurations are the leading ones. These patterns are generally inherited by the AcF molecule, resulting in a very dense spectrum of electronic states. For this reason, the electronic structure of AcF is strikingly different from that of TlF, which can be considered as an isovalent molecule. In contrast to Ac, one of the valence $s$-electrons ($6s$ electron in this case) of Tl never 
participates in low-energy excitations~\cite{Balasubramanian:85,Zou:09,Liu:TlF:20}.

The general feature of the majority of excited electronic states in AcF is their multireference character, meaning that no single determinant dominates the expansion of the electronic wave function. Therefore, the scheme used to calculate the IP based on the single-reference approximation cannot be applied to the excited states. The typical method of choice to solve such problems is the relativistic multireference Fock space coupled cluster theory (FS-RCC)~\cite{Kaldor:91,Visscher:01,Eliav:Review:22}. This approach is again based on the exponential parametrization of the many-electron wavefunction, but in contrast to Eq.~(\ref{expAnsatz}), exact wavefunctions $\Psi_i$ are obtained by acting on model vectors $\tilde{\Psi}_i$:
\begin{equation}
\ket{\Psi_i} = \{ e^T \} \ket{\tilde{\Psi}_i}, \quad\quad \ket{\tilde{\Psi}_i} = \sum_{m}C_{im} \ket{\Phi_m}.
\label{eq:fscc-ansatz}
\end{equation}
The model vectors represent the leading terms in the expansions of full electronic wavefunctions and are constructed from the model-space Slater determinants $\Phi_m$. The two-particle Fock-space sector ($0h2p$) determinants $\Phi_m$ used in the present paper are obtained from the Fermi vacuum determinant $\Phi_0$ (the $0h0p$ sector) by adding two electrons and distributing them in chosen \emph{active} one-particle spinors in all possible ways. The cluster operator $\hat{T}$ is also extended to enable electronic excitations from these active spinors.

In order to be able to consider very large model spaces required for a reasonable description of the multireference problem and to avoid convergence difficulties arising from the intruder-state problem that is typical for FS-RCC, formulations based on the intermediate Hamiltonian (IH) concept were developed (see \cite{Eliav:Review:22} and references therein). Among these formulations, the version that operates with incomplete main model spaces \cite{Zaitsevskii:QED:22} seems to be the most powerful. In this case, the whole model space for a given Fock-space sector is split into the main and auxiliary intermediate subspaces. The most accurate description that is stable with respect to the IH parameters is achieved for the states dominated by \emph{main} model subspace determinants (see Ref.~\cite{Zaitsevskii:QED:22} for the details of the IH formulation for the incomplete main-model space used here).

For AcF, the excited electronic states of interest can be best described within the IH-FS-RCC approach starting from the Fermi vacuum that corresponds to the closed-shell AcF$^{2+}$ cation. Note that the essentially two-electron nature of excitations from the closed-shell ground state prevents the use of FS-RCC in the one hole, one particle ($1h1p$) sector to describe the majority of states of AcF below 50,000~\cm. However, auxiliary IH-FS-RCCSD($1h1p$) calculations allowed us to conclude that charge-transfer states dominated by excitations from the $p$-shells of the F atom lie at energies $\sim$$72,000$~cm$^{-1}$, which is well above the IP value of 48,866~cm$^{-1}$ obtained in the present work.

\begin{figure}
    \centering
    \includegraphics[width=0.9\columnwidth]{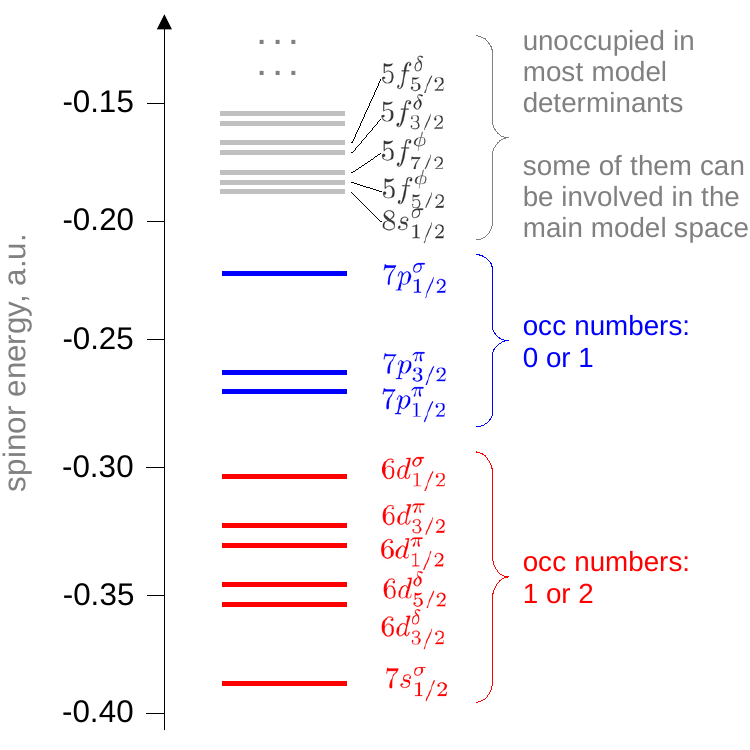}
    \caption{Energy diagram for active space of one-electron spinors of the AcF$^{2+}$ ion (the upper part is not shown). To define the incomplete main model space in the IH technique the active space is divided into three parts.}
    \label{fig:acf-spinors}
\end{figure}

The qualitative picture of the AcF electronic states that is described above also allows one to conveniently classify one-electron molecular spinors of AcF$^{2+}$ and to properly construct the main-model subspace for the IH-FS-RCC calculation. Overall, the 49 lowest-energy Kramers pairs of virtual spinors of AcF$^{2+}$ comprised the active space, which was further divided into three parts, as shown in Fig.~\ref{fig:acf-spinors}.

The first part comprised the lowest-lying $7s_{1/2}^\sigma$, $6d_{3/2}^\delta$, $6d_{5/2}^\delta$, $6d_{1/2}^\pi$, $6d_{3/2}^\pi$ and $6d_{1/2}^\sigma$ spinors. Here, the $7s_{1/2}^\sigma$ spinor refers to the molecular $\sigma_{1/2}$ spinor arising mainly from the $7s$ atomic orbital of Ac$^{2+}$, etc. Model-space determinants bearing one or two electrons on these spinors were considered as belonging to the main subspace. The second block of active spinors included $7p_{1/2}^\pi$, $7p_{3/2}^\pi$ and $7p_{1/2}^\sigma$ spinors, and the configurations with maximum one electron on them were also treated as the main ones. The maximum of zero-order energies of these ``main'' configurations was considered as the frontier energy which was then used to calculate the energy denominator shift parameters used in the IH-FS-RCC scheme (see Ref.~\cite{Zaitsevskii:QED:22} for technical details). Finally, all the remaining active spinors served to construct intermediate ``buffer'' determinants that were not included in the main model subspace. However, configurations with energies below the frontier energy, including $7s_{1/2}^\sigma8s_{1/2}^\sigma$ and $7s_{1/2}^\sigma5f$, were also added to the main subspace (despite the fact that the $8s_{1/2}^\sigma$ and $5f$ spinors belong to the third subspace in Fig.~\ref{fig:acf-spinors}) as they are necessary to maintain the correct physical picture. Such a composition of the incomplete main-model space allows us to reach nearly all electronic states below $\sim$43,000~cm$^{-1}$; states above this energy received significant contributions from ``buffer'' determinants and are thus expected to be described less accurately.

The IH-FS-RCCSD model with single and double excitations in the cluster operator was used for the excited-state calculations. To reduce systematic errors dependent on the internuclear separation and arising from the neglect of higher cluster amplitudes and basis set incompleteness, we combined excitation energies computed within the IH-FS-RCCSD model with the ground-state potential energy curve obtained at the CCSD(T) level (as it was previously done in Refs.~\cite{Isaev:2021,Zaitsevskii:2022,Zaitsevskii:ThO:23}). The counterpoise correction was applied to reduce the basis set superposition error. All molecular spinors up to the virtual energy cutoff of 300 Hartree were included in the correlation calculations.

Relativistic (including the Breit electron-electron interaction) and QED effects were effectively simulated using the generalized relativistic pseudopotential (GRPP) operator~\cite{Titov:99,Petrov:04b,Mosyagin:16,Zaitsevskii:QED:22}, whose high accuracy for actinide compounds has been previously demonstrated~\cite{Zaitsevskii:ThO:23,Oleynichenko:2023}. The new tiny-core GRPP for Ac, replacing 28 core electrons and accounting for the Breit interaction, finite-nuclear-size, and QED contributions (electron self-energy and vacuum polarization~\cite{Shabaev:13,Shabaev:18}), was constructed~\cite{GRPP_website}. Atomic low-lying excitation energies calculated for the neutral Ac atom and its Ac$^+$ cation at the relativistic two-component Hartree-Fock level with GRPP deviate from the reference all-electron Dirac-Coulomb-Breit + QED results by a few wavenumbers. These deviations do not exceed 20 cm$^{-1}$ even for the $5f$ states of Ac$^+$ (see Supplementary Materials), whereas the contributions from Breit interactions and QED effects reach 500 and 200 cm$^{-1}$, respectively. For the F atom, the empty-core GRPP that leaves all 9 electrons for explicit treatment and thus simulates only relativistic effects~\cite{Mosyagin:21} was employed. 

The basis set for Ac in the IH-FS-RCCSD calculations was based on the quadruple-zeta Dyall's basis set~\cite{Dyall:07,Dyall:12} and was comprised of the ($21s16p17d15f8g6h4i$) primitive Gaussian functions. For the F atom, the uncontracted version of the Dyall's AAEQZ basis set~\cite{Dyall:F:16} was employed. Such an extensive basis set ensures a high computational quality for a broad manifold of AcF electronic states, including those bearing the $8s$ and $5f$ electrons of Ac.

Transition dipole moments (TDMs) were calculated using the direct finite-order approach based on the substitution of the exponential wave operator into bra- and ket-vectors, and the infinite summation was truncated at terms quadratic in cluster amplitudes (\cite{Zaitsevskii:ThO:23}; see also Ref.~\cite{Blundell:91,Safronova:99,Gopakumar:02,Sahoo:05,Tan:23} for the analogous technique in atomic calculations). This approach allows one to obtain TDMs for all pairs of electronic states simultaneously. In contrast to the finite-field (FF) technique~\cite{Zaitsevskii:Optics:18,Zaitsevskii:TDM:2020}, it does not require any reduction in the molecular symmetry to calculate TDMs for the $\Delta\Omega=\pm 1$ case. To verify the accuracy of TDM calculations, the $0^+ - 0^+$ transition moments were also evaluated using the FF method for which the typical error is expected to be within few percent~\cite{Zaitsevskii:TDM:2020,Krumins:20,Kruzins:21}. Radiative decay rates and lifetimes of excited rovibrational states were evaluated using the Tellinghuisen’s sum rule~\cite{Tellinghuisen:84}.

Single-reference relativistic CCSD(T) calculations as well as solutions of the relativistic Hartree-Fock equations with subsequent integral transformation were obtained using the {\sc dirac} package~\cite{DIRAC_code:19,Saue:20} supplemented by the {\sc libgrpp} module~\cite{Oleynichenko:2023,LIBGRPP:23} to evaluate molecular integrals over generalized pseudopotentials. Excitation energies and TDMs were obtained within the {\sc exp-t} program package~\cite{EXPT_website,Oleynichenko_EXPT}. Molecular integrals over the operator of the projection of the electronic orbital angular momentum on the molecular axis $L_z$ that are required to calculate its expectation values were calculated using the OneProp code~\cite{Skripnikov:16b}. To solve the one-dimensional vibrational problem and calculate lifetimes, the {\sc vibrot} program~\cite{Sundholm} was used. Vibrational constants $\omega_e$ for each potential were derived from the first three calculated vibrational levels.

\section{Results and discussion}\label{sec:results}

\subsection*{Ionization potential and dissociation energy of AcF}

Contributions to the IP of AcF are given in Table~\ref{IPtable}. It is evident that correlation effects beyond the CCSD(T) model contribute quite significantly to the IP value, though they partly compensate for each other. 

Connected quadruple excitations that firstly appear in the CCSDT(Q) approach contribute about $+$166~\cm, while the CCSDTQ-3 model further corrects the IP value by $-$51~\cm. Contributions of the QED and Gaunt interaction effects were found to be quite small, $-$51 and $-$32 cm$^{-1}$, respectively. \textbf{ The final value for the IP$_e$ is 48,866(130)~cm$\bm{^{-1}}$}
(see Supplementary Materials for details of the employed scheme for the uncertainty estimation).

It should be noted that the IP$_e$ value was calculated as the difference of ground state electronic energies of the AcF$^+$ cation and AcF molecule. These energies were obtained in separate single-reference CC calculations with different Fermi-vacuum states. The precise experimental measurement of the IP will be able to probe the accuracy of this theoretical prediction, and it is thus valuable for testing the accuracy of the ground-state electronic wavefunctions of AcF$^+$ and AcF. As it was mentioned in the Introduction, the electronic ground state of AcF can be used as a working state to measure the Schiff moment of the Ac nucleus. An accurate interpretation of such an experiment will require the knowledge of the molecular enhancement constant that is determined by the ground-state electronic wavefunction of AcF. Although there is no explicit relation between IP$_e$ and the Schiff moment molecular enhancement, a good agreement between theoretical and experimental values for the IP is important to establish the reliability of electronic structure calculations.

\begin{table}[]
\caption{Calculated ionization potential and dissociation energy of AcF. The total correlation correction summarizes the contributions of correlation effects beyond the CCSD(T) model.}
\label{IPtable}
\renewcommand{\arraystretch}{1.2}
\begin{tabular*}{\columnwidth}{l@{\extracolsep{\fill}}rr}
\hline
\hline
Contribution            & IP, \cm   & $D$, \cm    \\
\hline
CCSD(T)                 & 49,083      & 57,209      \\
\\
~~CCSDT-3 $-$ CCSD(T)     & $-$246     & 63         \\
~~CCSDT $-$ CCSDT-3       & $-$93      & $-$347     \\
~~CCSDT(Q) $-$ CCSDT      & 166        & $-$24      \\
~~CCSDTQ-3 $-$ CCSDT(Q)   & $-$51      & 61         \\
Total correlation correction   & $-$224     & $-$247    \\
\\
Basis set correction    & 91         & 308        \\
QED                     & $-$51      & $-$24      \\
Gaunt                   & $-$32      & $-$32      \\
Zero-vibrational energy & 33         &  -269      \\
                        &            &            \\
Total, IP$_e$ and $D_e$    & 48,866      &  57,214      \\
Total, IP$_0$ and $D_0$    & 48,898      &  56,946      \\
\hline
\hline
\end{tabular*}
\end{table}

The upcoming experimental campaign by the CRIS collaboration~\cite{AcF:Proposal:21} also aims at an accurate determination of $D_e$ in AcF. Both the IP and $D_e$ are critical values for ionization processes during which dissociation can also occur, affecting the production techniques of actinium isotopes, including the medical radioisotope $^{225}$Ac. From the experimental data on the Ac ionization potential and F electron affinity~\cite{Sansonetti:05}, it follows that the ionic decay channel AcF~$\rightarrow$~Ac$^+$~($^1S_0$)~+~F$^-$~($^1S_0$) lies 15,962 cm$^{-1}$ above the neutral channel AcF~$\rightarrow$~Ac~($^2D_{3/2}$)~+~F~($^2P^o_{3/2}$). Therefore, the latter channel is considered. Obtained contributions to the dissociation energy of AcF are given in Table~\ref{IPtable}. One can see a slightly different contribution pattern compared to the IP case, though in both cases correlation effects beyond the CCSD(T) model have quite small contributions that partly compensate for each other. \textbf{The final value for the $\bm{D_e}$ is thus 57,214(234)~cm$\bm{^{-1}}$}.

\subsection*{Excited electronic states in AcF}

\begin{figure*}[ht]
    \centering
    \includegraphics[width=0.32\textwidth]{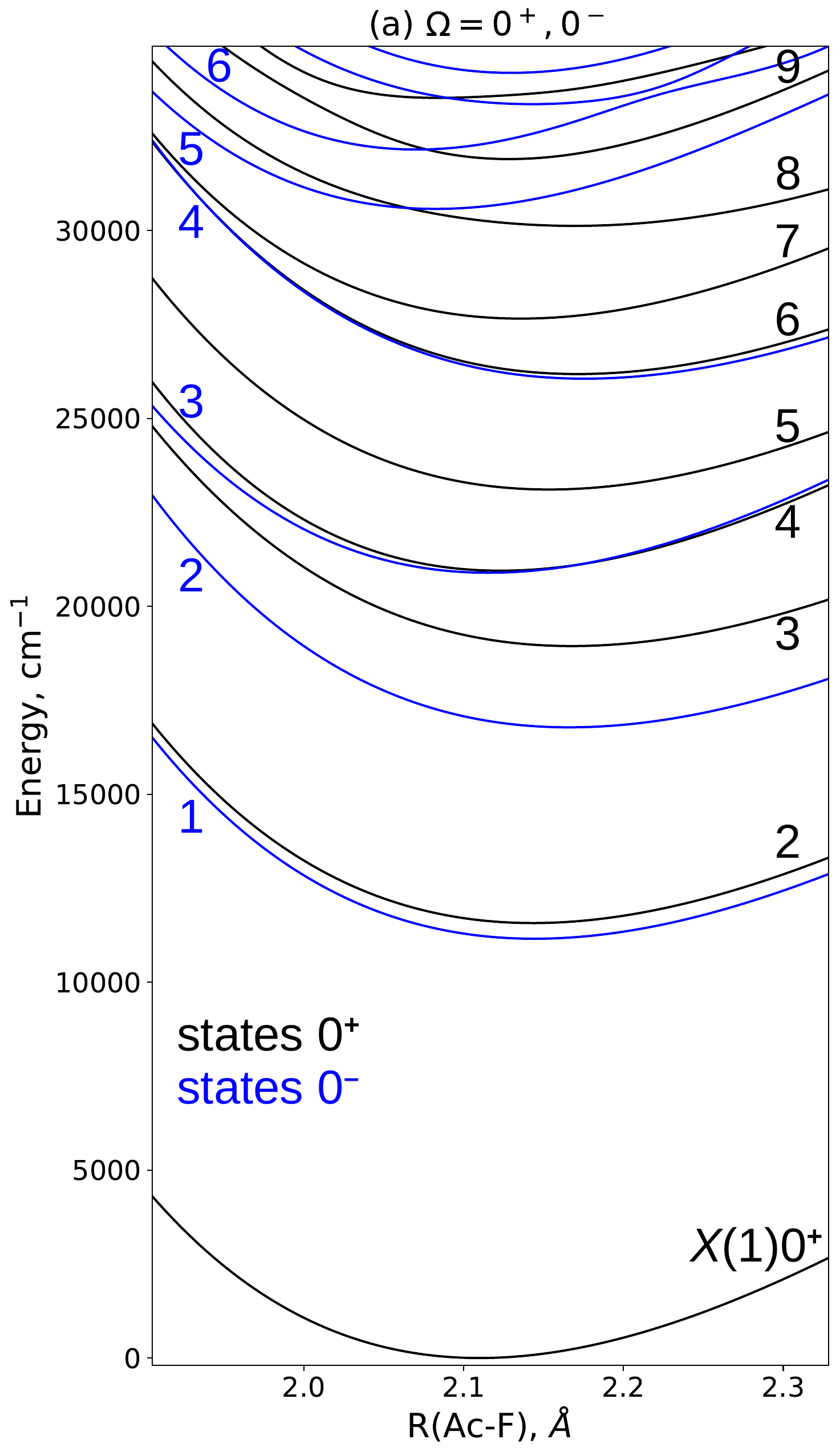}
    \includegraphics[width=0.32\textwidth]{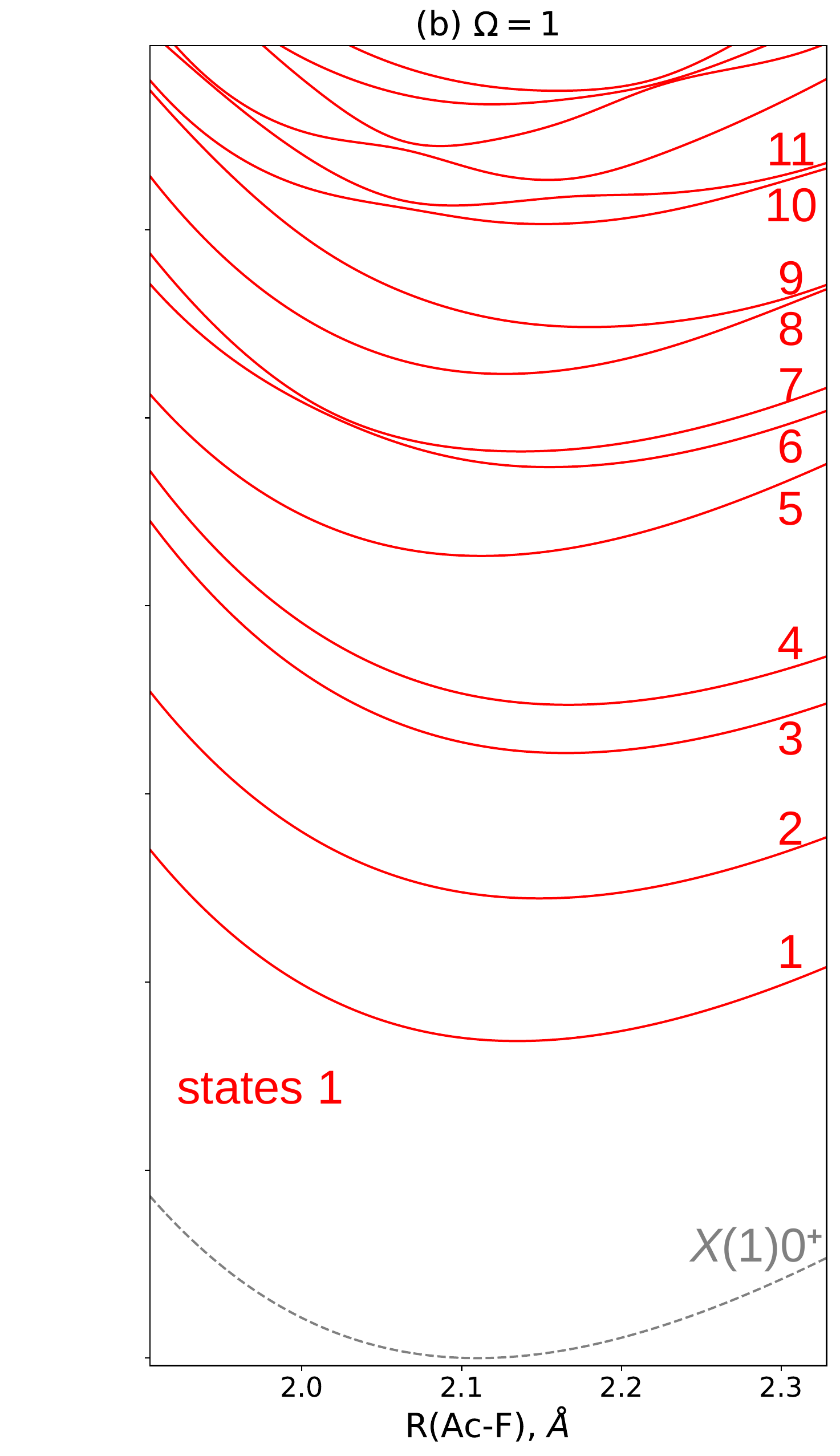}
    \includegraphics[width=0.32\textwidth]{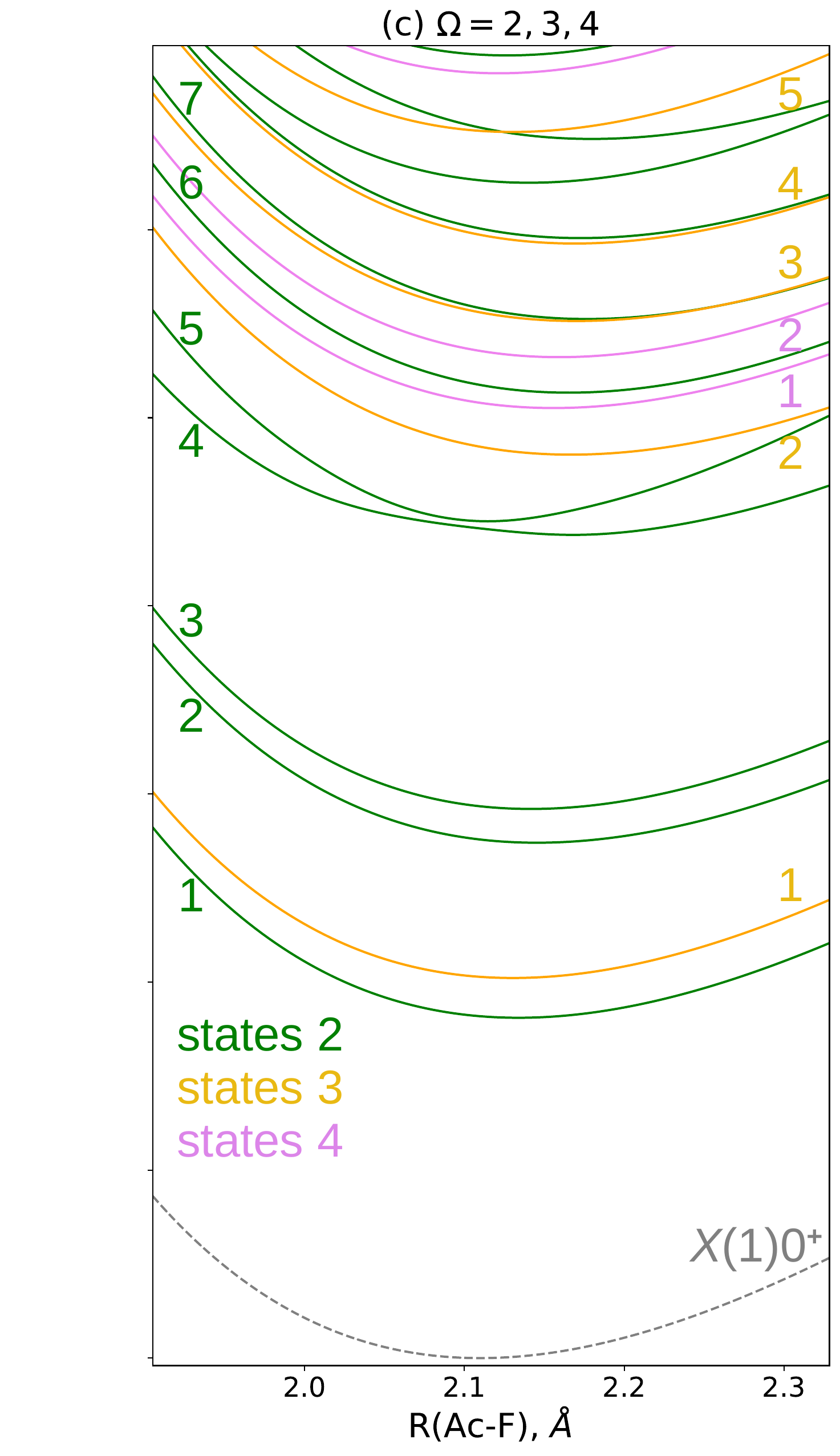}
    \caption{Potential energy curves of the AcF molecule in different electronic states: (a) $\Omega=0^+$, $0^-$; (b) $|\Omega|=1$; (c) $|\Omega|=$2, 3, 4. Energies are given with respect to the ground state equilibrium point.}
    \label{fig:pecs}
\end{figure*}

\begin{table*}[h]
\centering
\caption{Spectroscopic constants of $^{227}$AcF in different electronic states calculated at the IH-FS-RCCSD level. $|d|^2$ stands for the squared transition dipole moment for transitions from the $(1)0^+$ ground state. The values for $\braket{L_z}$, $|d|^2$, the compositions in terms of $\Lambda$S-states, and the leading configurations were calculated at $r(\text{\rm Ac-F}) = 4.0\ \text{a.u.} = 2.117$~\AA. $\omega_e$ values are absent for the states where the Born-Oppenheimer approximation is not applicable.}
\label{tab:spec-const}
\begin{tabular*}{\textwidth}{l@{\extracolsep{\fill}}ccccclll}
\hline
\hline
State & $T_e$, cm$^{-1}$ & $r_e$, \AA & $\omega_e$, cm$^{-1}$ & $|d|^2$, a.u. & $\braket{L_z}$, a.u. &  Composition & \multicolumn{2}{l}{Leading configurations} \\
\hline
 (1)$0^+$  &      0  &  2.110  &  541  &  --     &  0.0  & 100\% $X(1)^1\Sigma^+$ & 92\%  $7s_{1/2}^\sigma 7s_{1/2}^\sigma$ \\
 (2)$0^+$  &  11574  &  2.143  &  503  &  0.170  &  0.0  & 94\% $(1)^3\Pi$ & 88\%  $7s_{1/2}^\sigma 6d_{1/2}^\pi$ & 10\%  $7s_{1/2}^\sigma 7p_{1/2}^\pi$ \\
 (3)$0^+$  &  18944  &  2.168  &  479  &  2.145  &  0.0  & 78\% $(2)^1\Sigma^+$ + 15\% $(2)^3\Pi$ & 72\%  $7s_{1/2}^\sigma 6d_{1/2}^\sigma$ & 10\%  $6d_{3/2}^\delta 6d_{3/2}^\delta$ \\
 (4)$0^+$  &  20949  &  2.123  &  532  &  0.305  &  0.0  & 81\% $(2)^3\Pi$ + 14\% $(2)^1\Sigma^+$ & 76\%  $7s_{1/2}^\sigma 7p_{1/2}^\pi$ & \\
 (5)$0^+$  &  23109  &  2.153  &  494  &  0.286  &  0.0  & 75\% $(2)^3\Sigma^-$ + 12\% $(3)^1\Sigma^+$ & 54\%  $6d_{3/2}^\delta 6d_{3/2}^\delta$ & 16\%  $6d_{5/2}^\delta 6d_{5/2}^\delta$ \\
 (6)$0^+$  &  26182  &  2.171  &  479  &  0.000  &  0.0  &  90\% $(3)^3\Pi$ + 8\% $(2)^3\Sigma^-$ & 74\%  $6d_{3/2}^\delta 6d_{3/2}^\pi$ &  \\
 (7)$0^+$  &  27656  &  2.136  &  503  &  0.246  &  0.0  &  85\% $(3)^1\Sigma^+$ + 12\% $(2)^3\Sigma^-$ & 60\%  $6d_{5/2}^\delta 6d_{5/2}^\delta$ &  \\
 (8)$0^+$  &  30124  &  2.169  &  401  &  0.862  &  0.0  & 49\% $(4)^1\Sigma^+$ + 40\% $(4)^3\Sigma^-$ & 40\%  $6d_{1/2}^\pi 6d_{1/2}^\pi$ & 32\%  $7s_{1/2}^\sigma 7p_{1/2}^\sigma$ \\
 (9)$0^+$  &  31902  &  2.128  &  580  &  1.122  &  0.0  & 50\% $(4)^1\Sigma^+$ + 39\% $(4)^3\Sigma^-$ & 42\%  $7s_{1/2}^\sigma 7p_{1/2}^\sigma$ & 18\%  $6d_{3/2}^\pi 6d_{3/2}^\pi$ \\
(10)$0^+$  &  33529  &  2.084  &   --  &  0.002  &  0.0  & 52\% $(5)^1\Sigma^+$ + 40\% $(4)^3\Pi$ & 40\%  $6d_{3/2}^\delta 7p_{3/2}^\pi$ & 28\%  $7s_{1/2}^\sigma 8s_{1/2}^\sigma$ \\[1.5ex]
 (1)$0^-$  &  11156  &  2.144  &  502  &   --     &  0.0  & 85\% $(1)^3\Pi$ + 13\% $(1)^3\Sigma^-$ &  86\%  $7s_{1/2}^\sigma 6d_{1/2}^\pi$ &  \\
 (2)$0^-$  &  16781  &  2.166  &  487  &   --     &  0.0  & 82\% $(1)^3\Sigma^-$ + 14\% $(1)^3\Pi$ &  90\%  $7s_{1/2}^\sigma 6d_{1/2}^\sigma$ \\
 (3)$0^-$  &  20895  &  2.115  &  532  &   --     &  0.0  & 94\% $(2)^3\Pi$ &  86\%  $7s_{1/2}^\sigma 7p_{1/2}^\pi$  \\
 (4)$0^-$  &  26058  &  2.176  &  473  &   --     &  0.0  & 99\% $(3)^3\Pi$ &  80\%  $6d_{3/2}^\delta 6d_{3/2}^\pi$  \\
 (5)$0^-$  &  30575  &  2.082  &  540  &   --     &  0.0  & 98\% $(3)^3\Sigma^-$ &  84\%  $7s_{1/2}^\sigma 7p_{1/2}^\sigma$ \\
 (6)$0^-$  &  32155  &  2.070  &  589  &   --     &  0.0  & 97\% $(5)^3\Sigma^-$ &  84\%  $7s_{1/2}^\sigma 8s_{1/2}^\sigma$ \\
 (7)$0^-$  &  33363  &  2.143  &   --  &   --     &  0.0  & 72\% $(4)^3\Pi$ + 19\% $(6)^1\Sigma^+$ &  50\%  $6d_{3/2}^\delta 7p_{3/2}^\pi$ & 20\%  $6d_{1/2}^\sigma 6d_{1/2}^\pi$ \\
 (8)$0^-$  &  34196  &  2.130  &  559  &   --     &  0.0  & 33\% $(6)^1\Sigma^+$ + 25\% $(4)^3\Pi$ &  34\%  $6d_{1/2}^\pi    7p_{1/2}^\pi$ & 28\%  $6d_{3/2}^\delta 7p_{3/2}^\pi$ \\[1.5ex]
 (1)1      &   8430  &  2.135  &  514  &  0.001  &  2.0  & 99\% $(1)^3\Delta$ &  97\%  $7s_{1/2}^\sigma 6d_{3/2}^\delta$ \\
 (2)1      &  12222  &  2.149  &  499  &  0.244  &  0.9  & 78\% $(1)^3\Pi$ + 14\% $(1)^3\Sigma^-$ &  68\%  $7s_{1/2}^\sigma 6d_{1/2}^\pi$ & 16\%  $7s_{1/2}^\sigma 6d_{3/2}^\pi$ \\
 (3)1      &  16087  &  2.165  &  489  &  0.752  &  0.6  & 44\% $(1)^3\Sigma^-$ + 33\% $(1)^1\Pi$ &  49\%  $7s_{1/2}^\sigma 6d_{3/2}^\pi$    & 35\%  $7s_{1/2}^\sigma 6d_{1/2}^\sigma$ \\
 (4)1      &  17369  &  2.167  &  487  &  0.766  &  0.6  & 59\% $(1)^1\Pi$ + 39\% $(1)^3\Sigma^-$ &  53\%  $7s_{1/2}^\sigma  6d_{1/2}^\sigma$ & 18\%  $7s_{1/2}^\sigma  6d_{3/2}^\pi$ \\
 (5)1      &  21327  &  2.113  &  529  &  0.195  &  1.0  & 90\% $(2)^3\Pi$ + 6\% $(2)^1\Pi$ &  58\%  $7s_{1/2}^\sigma 7p_{1/2}^\pi$ & 22\%  $7s_{1/2}^\sigma 7p_{3/2}^\pi$ \\
 (6)1      &  23688  &  2.155  &   --  &  0.027  &  0.2  & 91\% $(2)^3\Sigma^-$ + 6\% $(3)^3\Pi$ &  78\%  $6d_{3/2}^\delta   6d_{5/2}^\delta$ \\
 (7)1      &  24105  &  2.137  &   --  &  1.706  &  1.1  & 49\% $(2)^1\Pi$ + 21\% $(3)^3\Pi$ &  31\%  $6d_{3/2}^\delta   6d_{1/2}^\pi$ & 29\%  $7s_{1/2}^\sigma   7p_{3/2}^\pi$ \\
 (8)1      &  26166  &  2.127  &  549  &  3.751  &  1.1  & 42\% $(2)^1\Pi$ + 30\% $(3)^3\Pi$ &  21\%  $7s_{1/2}^\sigma 7p_{3/2}^\pi$ & 19\%  $7s_{1/2}^\sigma 7p_{1/2}^\pi$ \\
 (9)1      &  27412  &  2.180  &  469  &  0.005  &  1.6  & 65\% $(2)^3\Delta$ + 31\% $(3)^3\Pi$ &  67\%  $6d_{3/2}^\delta   6d_{1/2}^\sigma$ &  \\
(10)1      &  30150  &  2.151  &   --  &  0.815  &  1.0  & 87\% $(3)^1\Pi$ + 8\% $(3)^3\Pi$ &  48\%   $6d_{5/2}^\delta  6d_{3/2}^\pi$ & 9\%   $6d_{3/2}^\delta  6d_{1/2}^\pi$ \\
(11)1      &  30643  &  2.094  &   --  &  0.120  &  0.1  & 89\% $(3)^3\Sigma^-$ + 6\% $(4)^3\Sigma^-$ &  77\%  $7s_{1/2}^\sigma   7p_{1/2}^\sigma$ \\
(12)1      &  31324  &  2.154  &   --  &  0.001  &  0.1  & 86\% $(4)^3\Sigma^-$ + 7\% $(3)^3\Sigma^-$ &  69\%   $6d_{1/2}^\pi      6d_{3/2}^\pi$ \\
(13)1      &  32224  &  2.087  &   --  &  0.019  &  0.0  & 98\% $(5)^3\Sigma^-$ &  84\%  $7s_{1/2}^\sigma   8s_{1/2}^\sigma$ \\
(14)1      &  33334  &  2.119  &   --  &  0.218  &  1.0  & 66\% $(4)^3\Pi$ + 29\% $(4)^1\Pi$ &  62\%  $6d_{3/2}^\delta   7p_{1/2}^\pi$ \\
(15)1      &  33694  &  2.160  &   --  &  0.001  &  1.8  & 81\% $(3)^3\Delta$ +8\% $(5)^3\Pi$ &  59\%  $6d_{1/2}^\pi      7p_{1/2}^\pi$ & 20\%   $6d_{1/2}^\pi      6d_{1/2}^\sigma$ \\[1.5ex]
 (1)2      &   9048  &  2.134  &  514  &   --     &  2.0  & 96\% $(1)^3\Delta$ &  66\%  $7s_{1/2}^\sigma    6d_{3/2}^\delta$ & 31\%   $7s_{1/2}^\sigma    6d_{5/2}^\delta$ \\
 (2)2      &  13702  &  2.145  &  496  &   --     &  1.6  & 53\% $(1)^1\Delta$ + 42\% $(1)^3\Pi$ & 38\%  $7s_{1/2}^\sigma   6d_{5/2}^\delta$ & 37\%  $7s_{1/2}^\sigma   6d_{3/2}^\pi$ \\
 (3)2      &  14600  &  2.142  &  509  &   --     &  1.5  & 56\% $(1)^3\Pi$ + 43\% $(1)^1\Delta$ & 51\%  $7s_{1/2}^\sigma  6d_{3/2}^\pi$ & 22\%  $7s_{1/2}^\sigma  6d_{5/2}^\delta$ \\
 (4)2      &  21887  &  2.168  &   --  &   --     &  2.9  & 83\% $(1)^3\Phi$ + 6\% $(2)^3\Pi$ & 79\%  $6d_{3/2}^\delta   6d_{1/2}^\pi$  & 11\%  $6d_{3/2}^\delta   7p_{1/2}^\pi$ \\
 (5)2      &  22248  &  2.115  &   --  &   --     &  1.0  & 92\% $(2)^3\Pi$ + 6\% $(1)^3\Phi$ & 80\%  $7s_{1/2}^\sigma   7p_{3/2}^\pi$ \\
 (6)2      &  25668  &  2.165  &  490  &   --     &  1.2  & 79\% $(3)^3\Pi$ + 15\% $(2)^3\Delta$ & 73\%  $6d_{5/2}^\delta   6d_{1/2}^\pi$ \\
 (7)2      &  27626  &  2.177  &  475  &   --     &  2.0  & 55\% $(2)^3\Delta$ + 25\% $(2)^1\Delta$ & 73\%  $6d_{1/2}^\sigma   6d_{3/2}^\delta$ \\
 (8)2      &  29777  &  2.173  &  478  &   --     &  1.9  & 63\% $(2)^1\Delta$ + 21\% $(2)^3\Delta$ & 72\%  $6d_{5/2}^\delta    6d_{1/2}^\sigma$ \\
 (9)2      &  31249  &  2.140  &  505  &   --     &  2.9  & 95\% $(2)^3\Phi$ & 77\%  $6d_{3/2}^\delta  7p_{1/2}^\pi $ & 12\%  $6d_{3/2}^\delta  6d_{1/2}^\pi $ \\
(10)2      &  32413  &  2.180  &  463  &   --     &  1.9  & 84\% $(3)^1\Delta$ + 7\% $(5)^3\Pi$ & 55\%  $6d_{3/2}^\pi      6d_{1/2}^\pi$    & 11\%   $6d_{3/2}^\pi      7p_{1/2}^\pi$ \\
(11)2      &  34634  &  2.127  &  515  &   --     &  1.1  & 93\% $(4)^3\Pi$ & 77\%  $6d_{5/2}^\delta   7p_{1/2}^\pi$ \\[1.5ex]
 (1)3      &  10106  &  2.130  &  518  &   --     &  2.0  & 100\% $(1)^3\Delta$ & 97\%  $7s_{1/2}^\sigma   6d_{5/2}^\delta$ \\
 (2)3      &  24020  &  2.167  &  481  &   --     &  2.9  & 91\% $(1)^3\Phi$ + 8\% $(2)^3\Delta$ & 51\%  $6d_{5/2}^\delta   6d_{1/2}^\pi$ & 31\% $6d_{3/2}^\delta   6d_{3/2}^\pi$ \\
 (3)3      &  27573  &  2.170  &  473  &   --     &  2.7  & 55\% $(1)^1\Phi$ + 38\% $(2)^3\Delta$ & 39\%  $6d_{3/2}^\delta   6d_{3/2}^\pi   $ & 25\%  $6d_{5/2}^\delta   6d_{1/2}^\sigma$ \\
 (4)3      &  29633  &  2.169  &  483  &   --     &  2.5  & 53\% $(2)^3\Delta$ + 44\% $(1)^1\Phi$ & 59\%  $6d_{5/2}^\delta   6d_{1/2}^\sigma$ & 11\%  $6d_{5/2}^\delta   6d_{1/2}^\pi   $ \\
 (5)3      &  32599  &  2.130  &  516  &   --     &  3.0  & 95\% $(2)^3\Phi$ & 49\%  $6d_{3/2}^\delta   7p_{3/2}^\pi$ & 28\%  $6d_{5/2}^\delta   7p_{1/2}^\pi$ \\[1.5ex]
 (1)4      &  25259  &  2.156  &  485  &   --     &  3.8  & 76\% $(1)^1\Gamma$ + 23\% $(1)^3\Phi$ & 76\%  $6d_{5/2}^\delta   6d_{3/2}^\delta$ & 14\%  $6d_{5/2}^\delta   6d_{3/2}^\pi   $ \\
 (2)4      &  26612  &  2.159  &  495  &   --     &  3.2  & 76\% $(1)^3\Phi$ + 24\% $(1)^1\Gamma$ & 68\%  $6d_{5/2}^\delta    6d_{3/2}^\pi   $ & 15\%  $6d_{5/2}^\delta    6d_{3/2}^\delta$ \\
 (3)4      &  34159  &  2.122  &  525  &   --     &  3.0  & 99\% $(2)^3\Phi$ & 77\%  $6d_{5/2}^\delta    7p_{3/2}^\pi $ & 12\%  $6d_{5/2}^\delta    6d_{3/2}^\pi $ \\
\hline
\hline
\end{tabular*}
\end{table*}

The potential energy curves of the AcF molecule shown in Fig.~\ref{fig:pecs} were calculated for the range of internuclear separations from 3.60~a.u. (1.905~\AA) to 4.40~a.u. (2.328~\AA). The chosen range is large enough to ensure the accurate numerical evaluation of wavefunctions and energies of at least the three lowest vibrational levels for each of the considered electronic states. The energy range up to $\sim$43,000~cm$^{-1}$ is limited by the composition of the main model space adopted in the present calculation. Spectroscopic constants including term energies $T_e$, equilibrium internuclear distances $r_e$, and vibrational constants $\omega_e$ for the states below 35,000 cm$^{-1}$ are summarized in Table~\ref{tab:spec-const}.

The IH-FS-RCCSD calculations confirm previously published predictions~\cite{Skripnikov:2020c,AcF:Proposal:21} that the electronic ground state is a closed-shell singlet with $\Omega = 0^+$. Potential energy curves for most of the considered electronic states are nearly parallel to each other, with $r_e$ lying well between 2.1 and 2.2~\AA, indicating that the electronic excitations are localized at the Ac atom as expected. The general pattern of potential energy curves closely resembles those of other actinide-containing diatomic molecules (for example, ThO~\cite{Zaitsevskii:ThO:23}, ThF$^+$~\cite{Denis:ThF:15}, and PaF$^{3+}$~\cite{Zulch:22}). The density of electronic states rapidly grows for the energy range above $\sim$30,000 cm$^{-1}$, resulting in a very complicated picture with many avoided crossings. This energy range does not seem to be convenient for experimental studies, since a firm interpretation of observed transitions would be hardly possible without the construction of comprehensive non-adiabatic rovibronic models~\cite{LefebvreBrion:04,Pazyuk:15}, which in turn require a large amount of spectroscopic data to be reliable. For the same reason, one can expect a substantially perturbed vibronic spectrum for transitions involving the $(6)1 \sim (7)1$ and $(4)2 \sim (5)2$ complexes (see Fig.~\ref{fig:pecs}b and~\ref{fig:pecs}c).

To better understand the structure of AcF, it is useful to represent the relativistic molecular electronic states in terms of their scalar-relativistic ($\Lambda - S$) counterparts. To perform such an analysis, an additional IH-FS-RCCSD calculation with the spin-orbit parts of pseudopotentials nearly switched off was performed for the point $r = 4.0$~a.u.~$= 2.117$~\AA{} that is close to the equilibrium distance in the majority of the studied AcF electronic states. The model-space parts of the obtained scalar relativistic states were then projected onto the manifold of fully relativistic states (see~\cite{Zaitsevskii:RbCs:17} for details). The results are summarized in Tab.~\ref{tab:spec-const}. For the majority of states presented in Tab.~\ref{tab:spec-const}, compositions in terms of $\Lambda-S$ states are far from pure, which is also illustrated by the expectation values of the $L_z$ operator. This observation clearly emphasizes 
the important
role of the spin-orbit interaction in this system.

Such an analysis also sheds light onto the nature of the $(6)1\sim(7)1$ and $(4)2\sim(5)2$ complexes. The avoided crossing in the former (see Fig.~\ref{fig:pecs}b) arises from the spin-orbit interaction between the $(2)^3\Sigma^-$ and $(2)^1\Pi$ components that dominate these states and is thus characteristic of a purely relativistic picture. The situation is similar for the $(4)2\sim(5)2$ complex (see Fig.~\ref{fig:pecs}c), since these two states are strongly dominated by the $(1)^3\Phi$ and $(2)^3\Pi$ scalar-relativistic states, respectively. In both of these cases, the crossing points lie close to $r_e$, resulting in the necessity to construct non-adiabatic models to treat their rovibrational structure. Such models are typically constructed with the involvement of a large amount of experimental data~\cite{LefebvreBrion:04,Pazyuk:15}. This seems to be unrealistic at the present level of theory. The avoided crossing also exists between the $(8)1$ and $(9)1$ states at $r \approx 2.3$~\AA, but it is expected that few unperturbed vibrational states exist for the $(8)1$ potential. In this case, the avoided crossing also stems from the spin-orbit coupling, mainly between the $(2)^3\Delta$ and the mixture of the $(2)^1\Pi$ and $(3)^3\Pi$ states.

It is important to estimate the typical uncertainty of the term energies $T_e$ given in Table~\ref{tab:spec-const}. The suggested empirical rule is that for FS-RCCSD calculations, the error in the ionization potential hardly exceeds the errors in excitation energies. For AcF, the IH-FS-RCCSD value of the IP equals to 49,170~cm$^{-1}$, which is 324~cm$^{-1}$ higher than the more accurate value obtained within the single-reference approach including triple and quadruple excitations (48,866~cm$^{-1}$). Taking into account the uncertainty of the latter (130~cm$^{-1}$), the uncertainty estimate for the excitation energies is $\sim$450~cm$^{-1}$. This value agrees well with the recently reported errors of IH-FS-RCCSD for the ThO molecule (which has a fairly similar electronic structure) at $<$~400~cm$^{-1}$ with a root-mean-squared deviation 280~cm$^{-1}$~\cite{Zaitsevskii:ThO:23}. The bulk of this uncertainty is expected to be due to the absence of triple excitations in the cluster operator. It is worth noting that the uncertainty is comparable to $\omega_e$ values, thus the present accuracy can be insufficient for an unambiguous assignment of experimentally measured vibrational bands.

\subsection*{Radiative properties of ground and excited states}

\begin{figure}[ht]
    \centering
    \includegraphics[width=\columnwidth]{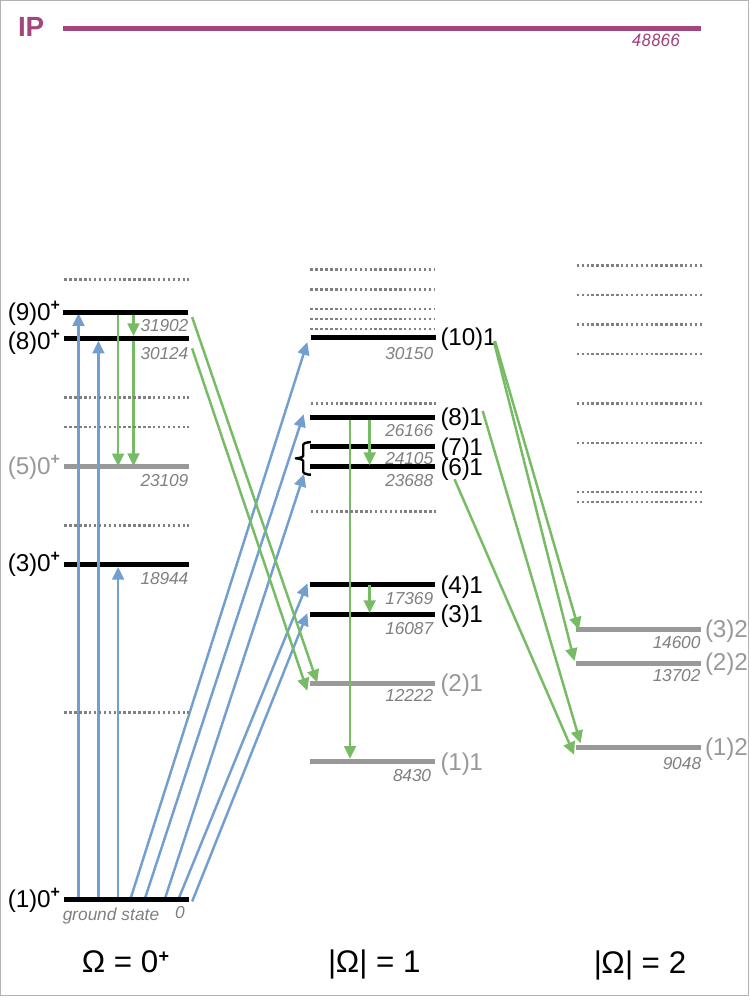}
    \caption{The strongest transitions (blue arrows) from the $X(1)0^+$ ground state of AcF and the strongest transitions for stimulated emission (green arrows). Levels accessible with two-step excitations are shown with solid grey lines. Dotted lines depict electronic states that are hardly accessible from the ground state with either direct or two-step excitations. It is noted that all transitions to the $\Omega=0^-$ states have low probabilities and are not shown here. $T_e$ values (cm$^{-1}$) are shown.}
    \label{fig:trans-scheme}
\end{figure}

\begin{figure}[ht]
    \centering
    \includegraphics[width=\columnwidth]{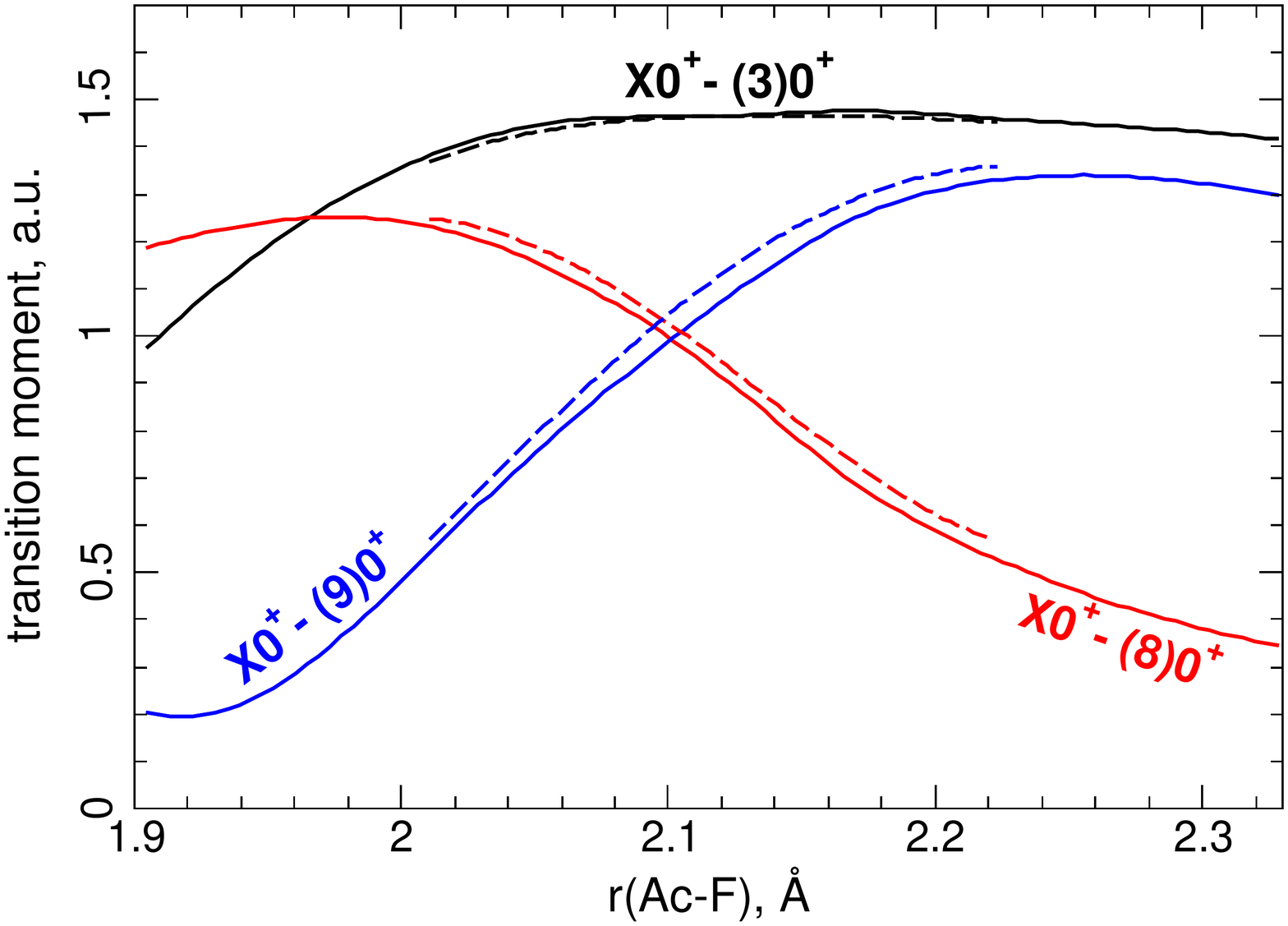}
    \caption{Transition dipole moment functions for the most intense transitions from the $X0^+$ ground state to excited states with $\Omega=0^+$. Solid and dashed lines denote the direct~\cite{Zaitsevskii:ThO:23} and finite-field~\cite{Zaitsevskii:Optics:18} methods, respectively.}
    \label{fig:tdm_X_0+}
\end{figure}

\begin{figure}[ht]
    \centering
    \includegraphics[width=\columnwidth]{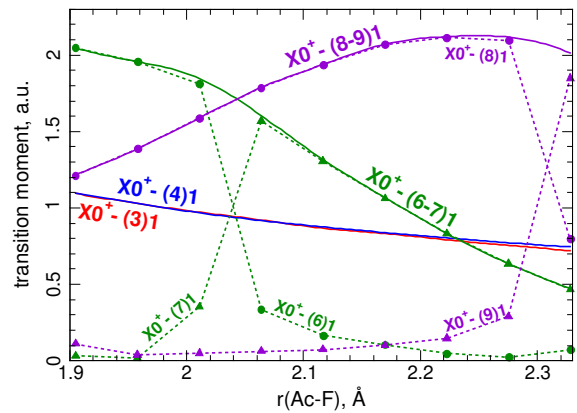}
    \caption{Transition dipole moment (TDM) functions for the most intense transitions from the $X0^+$ ground state to excited states with $|\Omega|=1$. Abrupt changes in TDM functions are due to the presence of avoided crossings. For the $(6)1 \sim (7)1$ and $(8)1 \sim (9)1$ complexes, a transformation from adiabatic to spectroscopic ``quasidiabatic'' states was performed (see Refs.~\cite{Krumins:20,Kruzins:21} for details).}
    \label{fig:tdm_X_1}
\end{figure}

The planned CRIS experiment on AcF at ISOLDE operates with radioactive ion beams at very low intensity (due to a typical production rate of $\sim$$10^6$ molecules/s or lower~\cite{Au2023}). Therefore, the identification of the most intense transitions is crucial. The probability of absorption of a photon in the $E1$ approximation is proportional to the square of the TDM $|d|^2$. Since the potential energy curves of excited states are roughly parallel to that of the ground state, and $r_e$ changes only slightly during the excitation, it is natural to compare intensities of vertical transitions (see Table~\ref{tab:spec-const}). The scheme in Fig.~\ref{fig:trans-scheme} depicts the strongest transitions from the ground state and secondary stimulated emission transitions (discussed below). Based on the obtained data, we expect only three intense transitions to $\Omega = 0^+$ states, $(3)0^+$, $(8)0^+$, $(9)0^+$, and five transitions to $|\Omega| = 1$ states, (3)1, (4)1, (6)1$\sim$(7)1, (8)1, (10)1. Clear unperturbed vibrational progressions could be expected for all these states except for the (6)1$\sim$(7)1 spin-orbit-coupled complex and the (10)1 state that belongs to the complex of several states at $\sim$30,000 -- 34,000~cm$^{-1}$. TDM functions for the most intense transitions are given in Figs.~\ref{fig:tdm_X_0+} and~\ref{fig:tdm_X_1}.

For the $(6)1\sim(7)1$ and $(8)1\sim(9)1$ complexes, it seems more natural to pass from adiabatic states to their diabatic ``spectroscopic'' counterparts (see, for example, Refs.~\cite{Krumins:20,Kruzins:21,Zaitsevskii:ThO:23} for details), which can be approximately identified with the dominating $\Lambda-S$ states (see Tab.~\ref{tab:spec-const}). The exact diabatization can be rather cumbersome. However, in these two particular cases one can
use the assumption that the lowest $X(1)0^+$ state is approximately a pure singlet state. As a result, the quasi-diabatic $X0^+ - (2)^1\Pi$ TDM function can be obtained by the requirement of vanishing the formally spin-forbidden $X0^+ - (2)^3\Sigma^-$ and $X0^+ - (3)^3\Pi$ transitions; the same holds for the $(8)1\sim(9)1$ complex. This simple approach is not valid for transitions to these complexes from many other states. For this reason, we further estimate total probabilities of transitions to the whole complex from higher-lying states without diabatization. This does not distort the calculated partial lifetimes significantly if the gap between the components of the complex is much smaller than the transition energy.

\begin{table*}[h]
    \centering
    \caption{The most probable ($|d|^2 > 0.5$) \emph{vertical} transitions to \emph{higher-lying} states with energies below 43,000~cm$^{-1}$ from the excited states that are easily accessible from the ground state. Energy levels and $|d|^2$ values were calculated for $r(\text{\rm Ac-F}) = 4.0\ \text{a.u.} = 2.117$~\AA. $^\dagger$~denotes states obtained as the intermediate ones in the IH-FS-RCC technique.}
    \label{tab:tdm-transitions-up}
\renewcommand{\arraystretch}{1.2}
\begin{tabular*}{\textwidth}{l@{\extracolsep{\fill}}lcclcclcclcc}
\hline
\hline
Initial state &
\multicolumn{3}{c}{$\rightarrow\Omega=0^+$} &
\multicolumn{3}{c}{$\rightarrow\Omega=0^-$} &
\multicolumn{3}{c}{$\rightarrow|\Omega|=1$} & 
\multicolumn{3}{c}{$\rightarrow|\Omega|=2$} \\
\cmidrule{2-4} \cmidrule{5-7} \cmidrule{8-10} \cmidrule{11-13}
& state & $E$, cm$^{-1}$ & $|d|^2$,a.u.
& state & $E$, cm$^{-1}$ & $|d|^2$,a.u.
& state & $E$, cm$^{-1}$ & $|d|^2$,a.u.
& state & $E$, cm$^{-1}$ & $|d|^2$,a.u. \\
\hline
$(3)0^+\rightarrow$ &  $(4)0^+$   &  20948  &  0.71 & & & & (16)1  &  36024  &  0.59 \\
(19105 cm$^{-1}$)   &  $(15)0^+$  &  37414  &  0.52 & & & & (26)1  &  39602  &  1.06 \\
                    &  $(17)0^+$  &  39146  &  0.76 & & & & (32)1  &  42505  &  1.03 \\[2ex]
$(8)0^+\rightarrow$ &  $(9)0^+$   &  31910  &  1.23 & & & & (17)1  &  36268  &  0.54 \\
(30237 cm$^{-1}$)   &  $(10)0^+$  &  33570  &  0.78 & & & & (20)1  &  37602  &  0.88 \\
                    &  $(11)0^+$  &  33742  &  0.93 & & & & \\
                    &  $(17)0^+$  &  39146  &  1.62 & & & & \\[2ex]
$(9)0^+\rightarrow$ &  $(10)0^+$  &  33570  &  2.33 & & & & 18(1)  &  36900  &  0.87 \\
(31910 cm$^{-1}$)   &  $(11)0^+$  &  33742  &  2.28 & & & & 20(1)  &  37602  &  2.99 \\
                    &  $(17)0^+$  &  39146  &  5.87 & & & & 23(1)  &  38598  &  1.24 \\
                    &  $(18)0^+$  &  40032  &  0.54 & & & & \\[2ex]
(3)1$\rightarrow$   &  $(18)0^+$  &  40032  &  1.08 & $(9)0^-$   &  36136  &  0.75 & & & & (10)2  &  32652  &  0.98 \\
(16239 cm$^{-1}$)   &             &         &       & $(12)0^-$  &  38796  &  0.84 & & & & (15)2  &  38026  &  1.08 \\
                    &             &         &       & $(13)0^-$  &  38823  &  1.14 & & & & (17)2  &  39131  &  0.60 \\
                    &             &         &       & $(15)0^-$  &  40465$^\dagger$  &  0.86 & & & & (18)2  &  40302  &  0.56 \\
                    &             &         &       &            &         &       & & & & (21)2  &  41759  &  0.57 \\[2ex]
(4)1$\rightarrow$   & $(21)0^+$   &  42460  &  0.65 & $(7)0^-$   &  33399  &  0.80 & (23)1  &  38598  &  0.59 & (10)2  &  32652  &  0.52 \\
(17531 cm$^{-1}$)   &             &         &       & $(8)0^-$   &  34206  &  0.75 &        &         &       & (17)2  &  39131  &  1.16 \\
                    &             &         &       & $(12)0^-$  &  38796  &  1.88 &        &         &       & (21)2  &  41759  &  1.32 \\
                    &             &         &       & $(14)0^-$  &  39978  &  0.85 &        &         &       & (23)2  &  42476  &  0.57 \\[2ex]
(7)1$\rightarrow$   & $(11)0^+$  &  33742  &  0.68 & $(6)0^-$   &  32330  &  0.85 & & & & (13)2  &  37394  &  3.53 \\
(24125 cm$^{-1}$)   & $(13)0^+$  &  36395  &  0.62 & $(11)0^-$  &  37776  &  0.53 & & & & (17)2  &  39131  &  1.20 \\
                    & $(15)0^+$  &  37414  &  0.73 & $(12)0^-$  &  38796  &  0.91 & & & & \\
                    &            &         &       & $(13)0^-$  &  38823  &  0.58 & & & & \\[2ex]
(8)1$\rightarrow$   & $(10)0^+$  &  33570  &  1.16 & & & & (26)1  &  39602  &  0.69 & (13)2  &  37394  &  2.12 \\
(26171 cm$^{-1}$)   & $(11)0^+$  &  33742  &  1.49 & & & &        &         &       & (18)2  &  39131  &  0.73 \\
                    & $(13)0^+$  &  36395  &  1.39 & & & &        &         &       & (19)2  &  40302  &  0.96 \\
\hline
\hline
\end{tabular*}
\end{table*}

Additionally, it is found that the probabilities of direct one-photon transitions from the $X0^+$ state to the manifold of excited states above $\sim 35,000$~cm$^{-1}$ are rather small, with $|d|^2 < 0.1$~a.u. However, two-step excitation schemes from the ground state that use different intermediate states can be devised. Table~\ref{tab:tdm-transitions-up} summarizes the estimates of $|d|^2$ for such secondary vertical excitations. However, it must be noted that these data should be considered as quite approximate. The $|d|^2$ values should thus be interpreted as indicative of the presence of a transition to a state with a certain $|\Omega|$. Note that, to obtain the full probability of such a two-step excitation process, these $|d|^2$ values should be additionally multiplied by their counterparts from Tab.~\ref{tab:spec-const}. While the higher-lying states can thus be observed in principle, it is expected that due to the very complicated picture of excited states above 35,000~cm$^{-1}$, the experimentally measured vibrational progressions would be strongly perturbed and nearly uninformative.

\begin{table}[h]
    \centering
    \caption{Squared transition dipole moments $|d|^2$ for transitions to \emph{lower-lying states} and radiative lifetimes $\tau$ of several excited states with $\Omega=0^+$ of AcF accessible from the ground state. $|d|^2$ values were calculated for $r(\text{\rm Ac-F}) = 4.0\ \text{a.u.} = 2.117$~\AA. Partial lifetimes for the channels with negligible contributions to the total decay rates are not shown.}
    \label{tab:tdm-from-0+}
\renewcommand{\arraystretch}{1.2}
\begin{tabular*}{\columnwidth}{l@{\extracolsep{\fill}}cccccccc}
\hline
\hline
          & \multicolumn{2}{c}{(3)0$^+$ $\rightarrow$} & & \multicolumn{2}{c}{(8)0$^+$ $\rightarrow$} & & \multicolumn{2}{c}{(9)0$^+$ $\rightarrow$} \\
\cmidrule{2-3} \cmidrule{5-6} \cmidrule{8-9}
$\rightarrow$ & \multicolumn{1}{c}{$|d|^2$,a.u.} & \multicolumn{1}{c}{$\tau$}
          & & \multicolumn{1}{c}{$|d|^2$,a.u.} & \multicolumn{1}{c}{$\tau$}
          & & \multicolumn{1}{c}{$|d|^2$,a.u.} & \multicolumn{1}{c}{$\tau$} \\
\hline
(1)0$^+$  &  2.145  & 35.2 ns     & & 0.862  & 34.8 ns     & & 1.122 & 12.3 ns      \\
(2)0$^+$  &  0.050  & 33.3 $\mu$s & & 0.097  & 736 ns      & & 0.077 & 836 ns       \\
(3)0$^+$  &         &             & & 0.246  & 1.63 $\mu$s & & 0.077 & 2.34 $\mu$s  \\
(4)0$^+$  &         &             & & 0.001  & 624 $\mu$s  & & 0.095 & 4.21 $\mu$s  \\
(5)0$^+$  &         &             & & 0.474  & 2.55 $\mu$s & & 0.433 & 1.85 $\mu$s  \\
(6)0$^+$  &         &             & & 0.000  & 44.9 ms     & & 0.008 & 362 $\mu$s   \\
(7)0$^+$  &         &             & & 0.229  & 201 $\mu$s  & & 0.066 & 72.6 $\mu$s  \\
(8)0$^+$  &         &             & &        &             & & 1.234 & 83.0 $\mu$s  \\[1ex]
 (1)1     &  0.062  & 8.89 $\mu$s & & 0.024  & 1.88 $\mu$s & & 0.007 & 5.63 $\mu$s  \\
 (2)1     &  0.001  & 672  $\mu$s & & 0.810  & 92.9 ns     & & 0.752 & 98.5 ns  \\
 (3)1     &  0.029  & 999  $\mu$s & & 0.150  & 1.10 $\mu$s & & 0.008 & 17.8 $\mu$s  \\
 (4)1     &  0.011  & 16.5 ms     & & 0.011  & 63.0 $\mu$s & & 0.109 & 1.59 $\mu$s  \\
 (5)1     &         &             & & 0.335  & 1.94 $\mu$s & & 0.232 & 1.83 $\mu$s  \\
(6-7)1    &         &             & & 0.203  & 10.0 $\mu$s & & 0.037 & 38.1 $\mu$s  \\
 (8)1     &         &             & & 0.301  & 40.1 $\mu$s & & 0.378 & 7.30 $\mu$s  \\
 (9)1     &         &             & & 0.133  & 171  $\mu$s & & 0.047 & 168 $\mu$s  \\
(10)1     &         &             & & 0.131  &             & & 0.001  \\
(11)1     &         &             & &        &             & & 0.016  \\
(12)1     &         &             & &        &             & & 0.000  \\
\\
$\tau_{\mathrm{total}}$ & & 35.0 ns & & & 22.8 ns &  & & 10.5 ns \\
\hline
\hline
\end{tabular*}
\end{table}

\begin{table}[h]
    \centering
    \caption{Squared transition dipole moments $|d|^2$ for transitions to \emph{lower-lying} states and radiative lifetimes $\tau$ of several excited states with $|\Omega|=1$ of AcF accessible from the ground state. $|d|^2$ values were calculated for $r(\text{\rm Ac-F}) = 4.0\ \text{a.u.} = 2.117$~\AA. Partial lifetimes for the channels with negligible contributions to the total decay rates are not shown.}
    \label{tab:tdm-from-1}
\renewcommand{\arraystretch}{1.2}
\begin{tabular*}{\columnwidth}{l@{\extracolsep{\fill}}cccccccc}
\hline
\hline
          & \multicolumn{2}{c}{(3)1 $\rightarrow$} & & \multicolumn{2}{c}{(4)1 $\rightarrow$} & & \multicolumn{2}{c}{(8)1 $\rightarrow$} \\
\cmidrule{2-3} \cmidrule{5-6} \cmidrule{8-9}
$\rightarrow$ & \multicolumn{1}{c}{$|d|^2$,a.u.} & \multicolumn{1}{c}{$\tau$}
          & & \multicolumn{1}{c}{$|d|^2$,a.u.} & \multicolumn{1}{c}{$\tau$}
          & & \multicolumn{1}{c}{$|d|^2$,a.u.} & \multicolumn{1}{c}{$\tau$} \\
\hline
(1)0$^+$  &  0.752  & 178 ns      & &  0.766  & 139 ns      & &  3.751 & 7.14 ns     \\
(2)0$^+$  &  0.071  & 77.0 $\mu$s & &  0.026  & 101 $\mu$s  & &  0.003 & 48.5 $\mu$s \\
(3)0$^+$  &         &             & &         &             & &  0.160 & 8.15 $\mu$s \\
(4)0$^+$  &         &             & &         &             & &  0.012 & 301 $\mu$s  \\
(5)0$^+$  &         &             & &         &             & &  0.099 & 211 $\mu$s  \\[1ex]
(1)0$^-$  &  0.072  & 57.3 $\mu$s & &  0.045  & 46.8 $\mu$s & &  0.004 & 40.3 $\mu$s \\
(2)0$^-$  &         &             & &  0.011  & 237 ms      & &  0.046 & 13.4 $\mu$s \\
(3)0$^-$  &         &             & &         &             & &  0.006 & 566 $\mu$s  \\[1ex]
(1)1      &  0.000  & 4.44 ms     & &  0.004  & 188 $\mu$s  & &  0.406 & 226 ns      \\
(2)1      &  0.005  & 1.27 ms     & &  0.088  & 39.1 $\mu$s & &  0.031 & 6.40 $\mu$s \\
(3)1      &         &             & &  0.421  & 574 $\mu$s  & &  0.022 & 17.8 $\mu$s \\
(4)1      &         &             & &         &             & &  0.152 & 4.70 $\mu$s \\
(5)1      &         &             & &         &             & &  0.049 & 92.9 $\mu$s \\
(6-7)1    &         &             & &         &             & &  0.615 & 93.7 $\mu$s \\[1ex]
(1)2      &  0.158  & 9.66 $\mu$s & &  0.059  & 15.4 $\mu$s & &  0.552 & 195 ns      \\
(2)2      &  0.002  & 1.44 ms     & &  0.168  & 42.8 $\mu$s & &  0.007 & 37.7 $\mu$s \\
(3)2      &  0.086  & 946 $\mu$s  & &  0.022  & 9.12 ms     & &  0.022 & 6.85 $\mu$s \\
(4-5)2    &         &             & &         &             & &  0.007 & 478 $\mu$s  \\
(6)2      &         &             & &         &             & &  0.003 &             \\[1ex]
$\tau_{\mathrm{total}}$ & & 174 ns & & & 136 ns &  & & 6.65 ns \\
\hline
\hline
\end{tabular*}
\end{table}

In contrast, stimulated emission to lower-lying excited states seems to be much more promising for experimental studies of the electronic spectrum of AcF. Tables~\ref{tab:tdm-from-0+} and~\ref{tab:tdm-from-1} present radiative properties of several excited states with $\Omega=0^+$ and $|\Omega|=1$, respectively. Based on these estimates, one can expect that at least four more excited states below 32,000~cm$^{-1}$ can be accessed and studied via such a two-step approach. The $(8)0^+$ and $(9)0^+$ states can be used to repump AcF to the $(5)0^+$ and $(2)1$ states, while the $(1)1$ and $(1)2$ components of the low-lying $(1)^3\Delta$ triplet are accessible via the $(8)1$ state. Note that the actual manifold of attainable electronic states is eventually determined by the capabilities of the experimental setup.

\subsection*{Suitability for direct laser cooling}

The fact that the potential energy curves are to some extent parallel to each other indicates that a closed optical loop might exist, which would make AcF suitable for direct laser cooling. To achieve laser coolability, the upper electronic state must meet three basic conditions~\cite{Isaev:Review:18,Tarbutt:18,Ivanov:19}: (a) a lifetime of order $10^1 - 10^2$ ns, (b) a quasidiagonal Franck-Condon matrix, and (c) absence of decays to other electronic states (or quite small branching ratios less than $10^{-3}$). It can be immediately seen from Tables~\ref{tab:tdm-from-0+} and~\ref{tab:tdm-from-1} that the last condition is definitely not fulfilled for the states $(8)0^+$ (the largest branching ratio $X0^+$:$(2)1$ equals to 1:0.38), $(9)0^+$ (1:0.13), $(3)1$ (1:0.02), $(4)1$ (1:0.009) and $(8)1$ (1:0.04).

The $(3)0^+$ state also possesses quite large but not fully prohibitve branching ratios, the largest of which (1:0.005) is for the $(3)0^+ \rightarrow (1)1$ decay channel. The sums for the first three (0.9876) and four (0.9980) Franck-Condon factors for the $X0^+ - (3)0^+$ transition are not optical, but do not allow one to unequivocally reject the existence of a quasi-closed optical loop. Thus, experimental data on the $X0^+$ and $(3)0^+$ states are necessary to resolve this question.

Finally we note that a closed loop employing some rovibronic state of the spin-orbit-coupled complex $(6)1 \sim (7)1$ is also possible in principle, but a non-adiabatic model that combines theoretical and experimental data is needed to clarify this question.

\subsection*{The ${(1)^3\Delta}$ state for $\mathcal{T}$,$\mathcal{P}$-violation searches}

\begin{figure}[ht]
    \centering
    \includegraphics[width=\columnwidth]{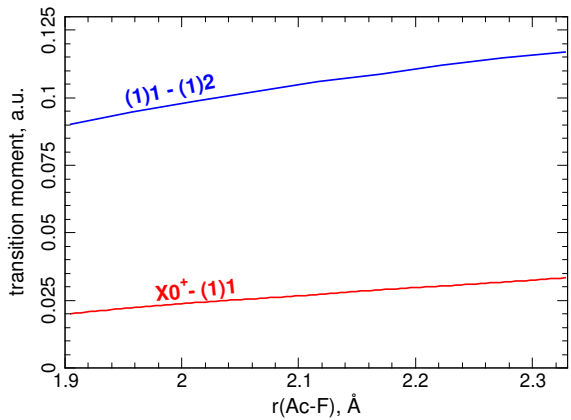}
    \caption{Transition dipole moment functions for transitions from the $(1)^3\Delta$ state components.}
    \label{fig:tdm_X_3Delta1}
\end{figure}

The three lowest-lying excited states of AcF, (1)1, (1)2 and (1)3 are components of the $(1)^3\Delta$ triplet state split by the spin-orbit interaction. In the non-relativistic realm and the $E1$ approximation, the $(1)^3\Delta \rightarrow X^1\Sigma^+$ transitions are forbidden, but the spin-orbit interaction opens the transition from the $^3\Delta_1 \equiv (1)1$ component to the ground state, making this state metastable. The same picture exists for the HfF$^+$ and ThO molecules~\cite{Petrov:09b,Leanhardt:2011,Skripnikov:16b,Tecmer:18,Zaitsevskii:ThO:23}, where the $^3\Delta_1$ metastable states were used to place the most stringent upper bound on the electron electric dipole moment ($e$EDM) to date~\cite{ACME:18,newlimit1}. TDMs between the $(1)^3\Delta_1$ and the ground electronic state and between the $|\Omega|=1$ and 2 components of $(1)^3\Delta$ are presented in Fig.~\ref{fig:tdm_X_3Delta1}. For the $(1)^3\Delta_1$ state, the sum rule~\cite{Tellinghuisen:84} gives a lifetime of 1.07 ms, which is approximately 4 times less than for ThO ($4.2(0.5)$ ms~\cite{Ang:22}) and 2000 times less than for HfF$^+$ ($2.1(0.2)$ s~\cite{Ni:HfF:14}). Thus, a shorter coherence time can be achieved with AcF than with systems currently used in $e$EDM experiments.

The lifetime of the $(1)^3\Delta_2$ state is predicted to be 181 ms (cf.~$> 62$~ms for ThO~\cite{Wu:ThO:20}). In principle, this electronic state could also be used to measure $\mathcal{T}$,$\mathcal{P}$-violating properties, such as the $e$EDM. However, the effective electric field acting on the $e$EDM in the $^3\Delta_2$ state is expected to be smaller than that for the $^3\Delta_1$ state. Additionally, the electronic $g$-factor in the $^3\Delta_2$ state is much larger than in the $^3\Delta_1$ state, where it is zero in the nonrelativistic approximation and thus allows for a significant reduction of systematic effects associated with stray magnetic fields.

\subsection*{The ground state for searches of nuclear properties}

The electronic ground state $X$~$^1\Sigma^+ \equiv X0^+$ of AcF can be used to search for the $\mathcal{T}$,$\mathcal{P}$-violating nuclear Schiff moment~\cite{Flambaum:Dzuba:20,Skripnikov:2020c}. The use of the ground state in such experiments implies that the coherence time is not limited by the lifetime of the working state. Moreover, other $\mathcal{T}$,$\mathcal{P}$-violating effects, e.g. those induced by the nuclear magnetic quadrupole moment, are strongly suppressed as they appear in the first order only for paramagnetic systems. Consequently, the interpretation of an experiment to measure the nuclear Schiff moment would be more direct.

Another possible use of the electronic ground state of AcF is the measurement of the Ac nuclear electric quadrupole moment. As it was demonstrated above, a highly accurate theoretical description of this state is tractable, and high-order coupled cluster methods are possible. Consequently, the electric field gradient at the Ac nucleus can be calculated with a very high accuracy. Therefore, a high-precision measurement of the molecular hyperfine structure can be used to extract the electric quadrupole moment of the actinium nucleus with high precision and accuracy.

\section{Conclusion}\label{sec:conc}

In the present work, the first comprehensive theoretical study of the electronic structure of AcF, with term energies below 35,000~cm$^{-1}$ and their radiative properties, is reported. We confirm the ground state to be a closed-shell singlet, and thus sensitive to the nuclear Schiff moment, and the IP$_e$ is calculated to be 48,866(130)~cm$^{-1}$. The obtained value for the dissociation energy is $D_e=57,214(234)$~\cm. The strongest transitions from the ground state are identified theoretically and can be used for experimental studies at the CRIS experiment at ISOLDE. Possible two-excitation pathways to excited electronic states below 43,000~cm$^{-1}$ that are not directly accessible from the ground one are proposed. Preliminary estimates based on calculated branching ratios and Franck-Condon factors allow one to assert that the molecule is not an ideal candidate for direct laser cooling, but experimental data on the $X0^+$ and $(3)0^+$ states are needed to unambiguously resolve the suitability. The lifetime of the metastable $(1)^3\Delta_1$ state, which can be used to search for the $\mathcal{T}$,$\mathcal{P}$-violating electric dipole moment of the electron
is, estimated to be 1.07 ms.

\section{Acknowledgements}

We are grateful to Anatoly V. Titov for useful discussions.
Electronic structure calculations have been carried out using computing resources of the federal collective usage center Complex for Simulation and Data Processing for Mega-science Facilities at National Research Centre ``Kurchatov Institute'', http://ckp.nrcki.ru/.

The computational study of AcF excited states and transition moments performed at NRC ``Kurchatov Institute'' -- PNPI by AVO, AZ and NSM was supported by the Russian Science Foundation (Grant No. 20-13-00225-P, https://rscf.ru/en/project/23-13-45028/). AcF ionization potential and dissociation energy calculations performed at NRC ``Kurchatov Institute'' -- PNPI by LVS were supported by the Russian Science Foundation Grant No. 19-72-10019-P (https://rscf.ru/en/project/22-72-41010/).
GN and MAK acknowledge support from the Belgian Excellence of Science project No. 40007501 (MANASLU) and the Flemish Science Foundation (FWO). MAU acknowledges support from the EU Horizon 2020 Research and Innovation Program No. 861198 (LISA) Marie Sklodowska-Curie Innovative Training Network (ITN).

\bibliography{JournAbbr,AcF,QCPNPI,SkripnikovLib,TitovLib,Maison,Prosnyak.bib}

\end{document}